\documentclass[journal=jctcce,manuscript=article]{achemso}
\usepackage[version=4]{mhchem}
\usepackage{amsmath,amssymb,bm}
\usepackage{booktabs}
\usepackage{array}
\usepackage{graphicx}
\usepackage{siunitx}
\author{Jun-qiang Lu}
\email{junqiang.lu@upr.edu}
\affiliation{Department of Physics, University of Puerto Rico at Mayag\"uez, Mayag\"uez, Puerto Rico 00680, USA}

\title{A Parameter-Free Reference Hamiltonian for Tight-Binding Vanadium Oxides}

\abbreviations{DFTB, PAW, SSF}

\begin{document}

\begin{tocentry}
\includegraphics{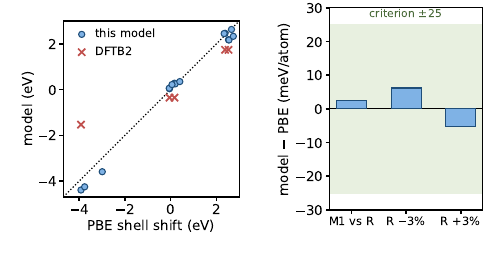}
\end{tocentry}

\begin{abstract}
Density-functional tight binding has yet to yield a transferable description of vanadium oxides; for the standard second-order construction and the tested cells, we demonstrate that two of its approximations deviate by electron volts and offset each other in the band-index centroid separation between the V~3d and O~2p bands: the neutral two-centre reference omits electron-volt environmental terms, whereas the Mulliken charge model assigns the compact V~3d shell less than half of the self-consistent potential shift observed in PBE.
Instead, we build a tight-binding model where the reference Hamiltonian derives from superposed neutral atoms, the full PBE exchange--correlation potential and frozen-core orthogonality, without adjustable terms, and where the orbital-resolved charge response employs three screening factors obtained from PBE linear response and an unscreened Coulomb range.
Relative to the primary PAW reference, the plane-wave Harris--Foulkes energy differences of \ce{VO2} are matched by the reference energy at fixed geometry and the stated numerical settings, with signed deviations of $+2.6$~meV/atom for M1 relative to rutile (an energy difference of \SI{92}{meV/atom}) and $+6.2$ and \SI{-5.2}{meV/atom} under $3\%$ volume compression and expansion; meanwhile, the charge response reproduces 14 of 22 fitted and 5 of 8 held-out matched PBE responses within one tolerance, and 28 of the 30 within twice the tolerance, and in rutile the fitted first-radial V~3d shift is \SI{-4.39}{eV}, \SI{0.47}{eV} more negative than PBE's \SI{-3.92}{eV}, whereas the tested DFTB2 charge model gives \SI{-1.53}{eV}.
The convexity of the spinless model at fixed geometry prevents it from spontaneously differentiating symmetry-equivalent sites under the symmetries respected by its implemented operators and sampling, so within this model such differentiation in the Magn\'eli phases must arise from electron--lattice coupling or an explicit spin or correlation term.
\end{abstract}

%======================================================================
\section{Introduction}
%======================================================================
One binary system of vanadium oxides hosts an unusual range of solid-state physics. As its vanadium atoms pair into dimers, \ce{VO2} transitions from a rutile metal to a monoclinic insulator near room temperature;\cite{Morin1959,Goodenough1971,Eyert2002} \ce{V2O3} exhibits its own metal--insulator transition; and the Magn\'eli series \ce{V_{n}O_{2n-1}} lies between them, forming crystallographic shear planes with coexisting \ce{V^{3+}}- and \ce{V^{4+}}-like sites.\cite{Andersson1954,Schwingenschlogl2004} Pinning down the localization of the extra electrons along a shear network and the dimerization of specific chains entails working with cells of $10^2$--$10^3$ atoms. Density-functional tight binding (DFTB) is designed to replace density-functional theory at this scale.\cite{Porezag1995,Elstner1998,Gaus2011,Hourahine2020}

Nevertheless, DFTB has not provided a transferable model for vanadium oxides. For the standard construction, we argue below that the limitation lies in two of its approximations rather than in the values of its parameters. Parameter sets covering entire elements that include vanadium, such as QUASINANO\cite{Wahiduzzaman2013} and the periodic-table baseline parameters for solids,\cite{Cui2024} choose the confinement within the same two-centre construction; GFN2-xTB\cite{Bannwarth2019} incorporates atomic multipole electrostatics but fits its Hamiltonian parameters. The standard second-order construction applies two approximations simultaneously. The reference Hamiltonian keeps only two-centre matrix elements of neutral atoms and sets free-atom eigenvalues on the diagonal. As a result, it leaves out the potential of third and further neighbours, the non-additivity of the exchange--correlation potential and the orthogonality of valence states to neighbouring cores. Transferred charge is represented in the charge response by Mulliken populations, sampled with Slater-type clouds whose widths are fixed by free-atom Hubbard parameters. Consequently, the induced potentials on the compact V~3d shell and the diffuse V~4s and 4p shells show a difference of about \SI{1}{eV}, whereas in PBE they differ by about \SI{4}{eV}. In vanadium oxides, both approximations err by electron volts, but their errors largely cancel in the separation between the V~3d and O~2p bands, as we show below. While a search over confinement radii produces band structures that appear close, it cannot correct both terms, since no radius modifies either approximation.

For the tested construction, cells and band-index centroid separation, we show that computing what DFTB approximates reduces the two compensating discrepancies and leaves only three screening factors to be determined, all from linear response. Defined as the full Kohn--Sham operator of superposed neutral PBE atoms, the reference Hamiltonian is projected out of the frozen cores and evaluated by real-space quadrature on declared, compactly supported atomic functions. No adjustable term is present. The reference energy is taken as the Harris--Foulkes energy at the superposed density,\cite{Harris1985,Foulkes1989} so no repulsive potential is fitted. The charge response maps the density-matrix change onto orbital-density clouds, while also screening the Coulomb kernel shell by shell. The three screening factors are determined by the matched PBE linear responses of \ce{VO2}-R and \ce{V2O3}, while the optimization selects the unscreened-range boundary $\eta = 0$.

The model reproduces the PAW Harris--Foulkes energy differences of \ce{VO2} at fixed geometry and the stated numerical settings, yielding signed deviations of $+2.59$~meV/atom for M1 relative to rutile and $+6.21$ and \SI{-5.20}{meV/atom} for $3\%$ volume compression and expansion (against the primary PAW reference). The model captures 14 of the 22 fitted responses and 5 of the 8 responses of the held-out M1 cell within one tolerance, and 28 of the 30 within twice the tolerance. At the level of \SI{0.03}{meV/atom}, the energy differences show no sensitivity to any tested refinement of the operator or the declared atomic functions; these are observed sensitivities, not bounds. The first-radial shell-shift fits deviate from PBE's by no more than \SI{0.6}{eV} across the three cells, including in M1, which was held out when we determined the response coefficients. The computed reference operator and the orbital-resolved response, with coefficients determined from linear-response data, give rise to these shell shifts, which were excluded from coefficient fitting.

%======================================================================
\section{Theory}
%======================================================================
\subsection{What DFTB approximates}
The Kohn--Sham energy expanded to second order about a reference density $\rho_0$ reads\cite{Foulkes1989,Elstner1998}
\begin{equation}
\begin{aligned}
E[P] &\simeq \mathrm{Tr}\big[P\,h[\rho_0]\big] + E_\mathrm{dc}[\rho_0] + \tfrac12 \iint \delta\rho(\mathbf r)\, K(\mathbf r,\mathbf r')\, \delta\rho(\mathbf r')\,\mathrm d\mathbf r\,\mathrm d\mathbf r' , \\
E_\mathrm{dc}[\rho_0] &= - E_\mathrm{H}[\rho_0] + E_\mathrm{xc}[\rho_0] - \int \rho_0\, v_\mathrm{xc}[\rho_0] + E_\mathrm{NN} ,
\end{aligned}
\label{eq:expansion}
\end{equation}
where $P$ is the one-particle density matrix, $\rho$ is its density and $\delta\rho = \rho - \rho_0$. Evaluating eq~\ref{eq:expansion} with the ground-state density matrix $P^0$ of $h[\rho_0]$ while omitting the second-order term yields the Harris--Foulkes functional,
\begin{equation}
E_\mathrm{HF}[\rho_0] = \mathrm{Tr}\big[P^0 h[\rho_0]\big] + E_\mathrm{dc}[\rho_0] .
\label{eq:harris}
\end{equation}
In this expression, $h[\rho_0] = T + v_\mathrm{ext} + v_\mathrm H[\rho_0] + v_\mathrm{xc}[\rho_0]$ and $K$ is the Hartree--exchange--correlation kernel.\cite{Harris1985,Foulkes1989} In second-order DFTB (DFTB2), each term is approximated:\cite{Elstner1998,Hourahine2020} specifically, $h[\rho_0]$ is replaced by two-centre integrals of confined atoms and free-atom eigenvalues on the diagonal, the double-counting terms and $E_\mathrm{NN}$ are replaced by a fitted pair repulsion, and the second-order term is replaced by $\tfrac12\sum_{AB}\Delta q_A\gamma_{AB}\Delta q_B$. In this context, $\Delta q$ represents Mulliken charges, while $\gamma$ constitutes a kernel derived from Slater-type clouds, with widths determined by atomic Hubbard parameters. The model developed below preserves the structure of eq~\ref{eq:expansion} while replacing each approximation.

\subsection{Reference operator}
We construct the reference density $\rho_0 = \sum_c \rho_c$ by superposing neutral, all-electron PBE atoms, including their cores, which were computed with a nominal confinement $(r/1000~\mathrm{bohr})^2$ having an energy of $4.4\times10^{-5}$~Ha for V and $1.2\times10^{-5}$~Ha for O, so that the atoms are unconfined in effect. We use V $3d^4 4s^1$ and O $2s^2 2p^4$ for the atomic configurations, following the PAW reference atom for V rather than the $3d^3 4s^2$ found in conventional DFTB parameterizations. The reference operator in the valence basis $\{\phi_\mu\}$ is
\begin{equation}
h^0_{\mu\nu} = \Big\langle Q\phi_\mu \Big| \, T + \sum_c v_{\mathrm{HN},c} + v_\mathrm{xc}^\mathrm{PBE}\Big[\sum_c\rho_c\Big] \Big| Q\phi_\nu \Big\rangle ,
\qquad Q = 1 - \sum_{a,b} |\chi_a\rangle (G^{-1})_{ab} \langle\chi_b| ,
\label{eq:h0}
\end{equation}
where $v_{\mathrm{HN},c}$ denotes the short-ranged nuclear plus Hartree potential of neutral atom $c$, the $\chi_a$ represent the frozen core orbitals (V 1s--3p, O 1s) of all atoms, and $G_{ab} = \langle\chi_a|\chi_b\rangle$ serves as their overlap matrix. Cores on neighbouring atoms overlap slightly, so this overlap matrix is not the unit matrix. We compute the exchange--correlation potential pointwise on the superposed density and its gradients, without ever splitting it into atomic pieces. The projection, employing the core--valence overlap $B = S_{cv}$ and $A = G^{-1}B$, produces the Bloch matrices
\begin{equation}
H^Q = H_{vv} - H_{vc}A - A^\dagger H_{cv} + A^\dagger H_{cc} A, \qquad S^Q = S_{vv} - B^\dagger A ,
\label{eq:qfold}
\end{equation}
which establish the generalized eigenvalue problem of eq~\ref{eq:h0} in the non-orthogonal basis. Prior to the Q fold, a scalar-relativistic on-site correction $d_\mu$ is added to the full core and valence blocks in the Hermitian form $\tfrac12(S\,\mathrm{diag}\,d + \mathrm{diag}\,d\,S)$. The shell's scalar-relativistic atomic eigenvalue minus the expectation value of the nonrelativistic atomic Hamiltonian, including the confinement potential for basis functions, in that orbital, evaluated on the original atomic radial before its tail continuation, equals this correction. The 3d value is carried by the two added d radials. No term in eqs~\ref{eq:h0}--\ref{eq:qfold} is adjusted.

Eigenfunctions of compressed atoms form the valence basis, employing a confinement potential $(r/r_{0,l})^2$ where $r_{0,l}$ = 6.38 (V~4s, 4p), 3.69 (V~3d) and 6.0~bohr (O~2s, 2p). The set includes O~2s and 2p, V~4s and 4p, plus three V~3d radials, namely the confined 3d function and its first and second derivatives with respect to the 3d occupation, taken along the $3d^4 4s^1$ path. We name this basis 3$\zeta$-d, while calculations with one and two d radials act as ablations.

\subsection{Declared atomic functions and quadrature}
For radii exceeding the matching radius $r_m$, we substitute every basis radial and free-atom density with its WKB continuation within the atom's own potential. A declared support marks where each continuation closes smoothly, defined by $r^2R^2$ dropping below $10^{-14}$ of its maximum for radials, or the density falling below $10^{-10}\,e/\mathrm{bohr}^3$ for densities. The WKB density of the highest occupied orbital forms the density tail, with a decay constant that approaches $\kappa_0 = \sqrt{-2\varepsilon_\mathrm{HOMO}}$ (V 0.534, O 0.815~bohr$^{-1}$). To ensure each atom remains exactly neutral, we locally restore the small amount of removed charge. By replacing the slowly decaying tails of the Slater-type expansions, this construction prevents the conditionally convergent Coulomb sums that would otherwise result. Details of the construction and its sensitivities are provided in the Supporting Information. We measure the effect of the density matching tolerance, the one registered quantity with an appreciable effect, on energy differences below.

Atom-centred real-space quadrature is employed to evaluate the reference-operator matrix elements, utilizing Stratmann--Scuseria--Frisch (SSF) partition weights with $a_\mathrm{SSF} = 0.64$,\cite{Stratmann1996} a rational radial map $r = r_\mathrm s\,u/(1-u)$ with $r_\mathrm s = 1$~bohr and 256 radial points, a $24\times48$ product angular grid, and common inclusion radii of 16.24 (V) and 10.95~bohr (O). A certificate verifies at each quadrature point that the periodic images contributing to its partition weight constitute a complete set. Prior to applying an operator, we verify at every $k$ point that its overlap matrix is Hermitian and contains no negative eigenvalues exceeding the guard's numerical tolerance.

\subsection{Reference energy}
The Harris--Foulkes energy at $\rho_0$, evaluated in the model basis,
\begin{equation}
E_\mathrm{ref} = \sum_{i\in\mathrm{val}} f_i\,\varepsilon_i[h^0] + \sum_{c} n_c\,\varepsilon_c^\mathrm{cryst} + E_\mathrm{NN} - E_\mathrm H[\rho_0] + E_\mathrm{xc}[\rho_0] - \int\rho_0\,v_\mathrm{xc}[\rho_0] \equiv C_0 + E_\mathrm{band} ,
\label{eq:eref}
\end{equation}
where $f_i$ and $\varepsilon_i[h^0]$ are the occupations and eigenvalues of the Q-projected valence Hamiltonian, constitutes the energy associated with $h^0$. The occupations of the frozen core levels are denoted by $n_c$, while $\varepsilon_c^\mathrm{cryst}$ represents their atomic energies plus the first-order shifts defined below. A point-charge sum, rather than a fitted term, constitutes the nuclear Coulomb repulsion $E_\mathrm{NN}$. Atomic constants are gathered in the geometry-dependent constant $C_0$, which also includes an electrostatic pair term between neutral atoms (tabulated per element pair with a range derived from the declared supports, 37~bohr), the exchange--correlation non-additivity $X[\rho_0]-\sum_A X[\rho_A]$ (where $X[\rho] = \int(\rho\,\varepsilon_\mathrm{xc} - \rho\,v_\rho - 2\sigma v_\sigma)$, $\sigma = |\nabla\rho|^2$, and $v_\rho$, $v_\sigma$ are the derivatives of $\rho\,\varepsilon_\mathrm{xc}$ with respect to $\rho$ and $\sigma$), and the first-order shifts of the frozen core levels in the crystal potential. Using Gaussian smearing (\SI{0.02}{eV}), we compute $E_\mathrm{band}$ and report the single-width energy estimate $E_{\sigma\to0} = C_0 + E_\mathrm{band} + \tfrac12(-TS)$. We fit no repulsive potential. Direct comparison with plane-wave Harris--Foulkes energies validates eq~\ref{eq:eref}.

\subsection{Charge response}
The density-matrix change $\Delta P = P - P_\mathrm{ref}$ is projected onto one cloud per shell and atom (V 4s, 4p, 3d; O 2s, 2p) using a charge-conserving scheme in the Coulomb metric. Occupations for the declared reference atoms, V $3d^4 4s^1$ and O $2s^2 2p^4$, are placed on the first radial of each shell within the reference density matrix $P_\mathrm{ref}$, ensuring that charges are measured from the same atomic configurations that define $\rho_0$; the added d radials carry none. The charges are given by $q_c = \sum_k w_k \mathrm{Tr}[L_c(k)\,\Delta P(k)]$. Since the clouds correspond to the orbital densities of the basis functions, the induced potential is sampled at the locations of each shell's electrons. Both the clouds and the mapping are constructed from products of basis functions prior to the frozen-core projection. The total induced cloud charge is zero, while the three d radials of an atom map onto its single 3d cloud. A mapping consistent with the Q projection, which encompasses both the adjoint and the reference subtraction, has not been validated. Core orbitals localized at the nuclei are the sole distinction between the projected and unprojected functions, yet this is not a bound on the resulting response error. The second-order energy and induced Hamiltonian are
\begin{equation}
\begin{aligned}
E_2 &= \tfrac12\, q^\mathrm T \Gamma q, \qquad \Delta h(k) = \sum_c (\Gamma q)_c\, L_c(k)^\dagger, \\
\Gamma &= D\,J\,D,\qquad J_{cc'} = \frac{1}{\Omega}\sum_{\mathbf G\neq0} \frac{4\pi}{|\mathbf G|^2+\eta}\, F_c^*(\mathbf G)F_{c'}(\mathbf G) ,
\end{aligned}
\label{eq:response}
\end{equation}
where $F_c$ denotes the Fourier transform of cloud $c$, $\Omega$ denotes the cell volume, and $D = \mathrm{diag}(s_c)$ imposes congruence screening via one factor for the V~3d clouds ($s_\mathrm{Vd}$), one for the V~4s/4p clouds ($s_\mathrm{Vsp}$) and one for the O clouds ($s_\mathrm O$). $\Gamma$ is positive semidefinite by construction, and no on-site counterterm is added. A nested hierarchy registered before the fit begins with the first level, comprising the three screening factors and the range coordinate, $\theta^{(0)} = (s_\mathrm{Vsp}, s_\mathrm{Vd}, s_\mathrm O, \eta)$. Additional shell factors are included only when the response fails along a resolvable direction.

Matched linear responses provide the basis for determining $\theta^{(0)}$. We impose a shell field of $\pm\SI{0.10}{eV}$ on all V~3d or all O~2p shells within a cell, adding a single V~3d shell test for \ce{VO2}-R. The occupation changes are recorded both without self-consistency (bare) and with it (screened). In PBE, the field acts via the PAW on-site occupation operator, mirroring the linear-response determination of Hubbard parameters.\cite{Cococcioni2005} Both responses are read using the same sphere projector $W_l$, constructed from the first all-electron partial wave of channel $l$ normalized inside a sphere of radius 2.30 (V) or 1.55~bohr (O). No scale is applied on either side, and the unscaled projector reproduces VASP's on-site occupation of the reference state to 91--93\% for V~3d and 97\% for O~2p. The field is applied as $\alpha W_l$ and read with $W_l$ in the model, whereas in PBE it is applied through VASP's on-site occupation operator and read with $W_l$. Only through this occupation has the correspondence of the two field operators been checked, which does not bound their difference.

Fermi--Dirac occupations of width $10^{-4}$~Ha are employed in the model's self-consistent and response calculations, distinct from the \SI{0.02}{eV} Gaussian convention applied to the reference energy. FFT grids of $72\times72\times48$ (rutile), $96\times72\times80$ (M1) and $84\times84\times84$ points (\ce{V2O3}) are used to represent the clouds and $J$. Only in the response calculations are the basis functions sampled on this grid truncated at the radius where $r^2R^2$ drops below $10^{-8}$ of its maximum (at most 15~bohr), whereas the cloud form factors integrate the full radial densities. A tolerance $\sigma = 0.10|\chi_\mathrm{PBE}| + \SI{0.005}{e/eV}$, $z = (\chi_\mathrm{model}-\chi_\mathrm{PBE})/\sigma$, is applied to compare each response.

%======================================================================
\section{Computational details}
%======================================================================
\begin{table}
\caption{The v0 model is defined by the parameters and numerical settings listed here. All lengths are given in bohr.}
\label{tab:params}
\begin{tabular}{@{}l>{\raggedright\arraybackslash}p{0.64\linewidth}@{}}
\toprule
item & value \\
\midrule
reference atoms & neutral, all-electron PBE, nominal $(r/1000)^2$ confinement; V $3d^4 4s^1$, O $2s^2 2p^4$ \\
frozen cores & V 1s--3p; O 1s \\
basis radials & O 2s, 2p; V 4s, 4p; V 3d$\times$3: $R_{3d}$, $\partial R_{3d}/\partial n_{3d}$, $\partial^2 R_{3d}/\partial n_{3d}^2$ along $3d^4 4s^1$ \\
basis confinement & $(r/r_{0,l})^2$ with $r_{0,l}$ = 6.3789 (V 4s, 4p), 3.6873 (V 3d), 6.0 (O 2s, 2p) \\
declared tails & WKB continuation beyond $r_m$; logarithmic-derivative tolerance 0.15; 1~bohr $C^1$ blend \\
declared supports & radials: $r^2R^2 < 10^{-14}$ of its maximum; densities: $\rho < 10^{-10}\,e/\mathrm{bohr}^3$ \\
density decay $\kappa_0$ & V 0.534, O 0.815~bohr$^{-1}$ \\
operator quadrature & SSF partition, $a_\mathrm{SSF} = 0.64$; $r = r_\mathrm s u/(1-u)$, $r_\mathrm s = 1$, 256 points; $24\times48$ angular \\
inclusion radii & V 16.24, O 10.95 \\
$C_0$ pair range & 37.0 \\
screening $\theta^{(0)}$ & $s_\mathrm{Vsp} = 0.597$, $s_\mathrm{Vd} = 0.633$, $s_\mathrm O = 0.934$, $\eta = 0$ \\
occupations & Fermi--Dirac, $10^{-4}$~Ha (self-consistency and response); Gaussian, \SI{0.02}{eV} ($E_\mathrm{ref}$) \\
\bottomrule
\end{tabular}
\end{table}

The parameters defining the v0 model are listed in Table~\ref{tab:params}. The Crystallography Open Database (COD)\cite{Grazulis2012} supplies the experimental structures employed here: \ce{VO2}-R (COD 1548819, $a = 4.5546$, $c = \SI{2.8528}{\angstrom}$ at \SI{373}{K}, deposited with ref~\citenum{Zhao2018}; 6 atoms), \ce{VO2}-M1 (COD 3000344, a room-temperature powder refinement with V and O sites only from the substitution series of ref~\citenum{Wilson2022}; 12 atoms, V--V distances 2.612 and \SI{3.172}{\angstrom}), and corundum \ce{V2O3} at ambient pressure (COD 9015569,\cite{Finger1980} 10-atom primitive cell). The model and PBE comparisons employ identical geometries, and the conclusions pertain to these selected structures. We scale the strained rutile cells isotropically in volume by $-3.008\%$ and $+2.991\%$.

All reference calculations employ the projector augmented-wave method\cite{Blochl1994,Kresse1999} within VASP\cite{Kresse1996} using the PBE\cite{Perdew1996} functional, a \SI{520}{eV} cutoff, \SI{0.02}{eV} Gaussian smearing, and no spin polarization (the reference-energy and band comparisons use the standard V potential with V~3p in the frozen core and the standard O potential, with V\_pv, 3p in the valence, as a second reference for the energies, whereas the response targets and the PBE calculations underlying the shell-shift and band-centroid analysis use V\_pv). Harris--Foulkes energies are computed non-self-consistently via superposed atomic charge densities (ICHARG = 12) on precisely the $k$-point sets adopted by the model (113 points for \ce{VO2}-R and the strained cells, 63 for M1); for \ce{V2O3}, a $\Gamma$-centred $5\times5\times5$ mesh is applied. PBE calculations for the same cells without spin polarization serve as the response targets. The LDAUTYPE = 3 machinery applies the shell field, with the density either held at its unperturbed self-consistent value (bare, ICHARG = 11) or relaxed (screened).

The quoted numbers are obtained under read-out conventions that were fixed in advance. The 3$\zeta$-d overlap matrices exhibit good conditioning, with the smallest eigenvalue of $S^Q$ being $3.5\times10^{-2}$ in \ce{VO2}-R and $1.9\times10^{-2}$ in M1. Consequently, we compute the candidate's energies at full rank. We evaluate differences between two cells at matched rank because the larger spans in the enrichment sequence below become nearly linearly dependent. At every $k$, we discard an identical count of weak overlap directions per formula unit in both cells, determining that count at each level from the overlap spectrum. We interpret this as a regularization of weak directions rather than the removal of null ones, given that the largest discarded eigenvalue is $1.5\times10^{-2}$. A distinct read-out, the absolute $S^Q$ eigenvalue cutoff $10^{-4}$, is employed for the band offsets.

We compute each matrix element of the reference operator through real-space integration over the specified quadrature. To avoid repeating assembly at every geometry, production workflows demand tabulating the calculated environment terms. We do not attempt this step in the present work.

%======================================================================
\section{Results}
%======================================================================
\subsection{Why DFTB2 fails for vanadium oxides, and what repairs it}
\begin{figure}
\centering
\includegraphics[width=\linewidth]{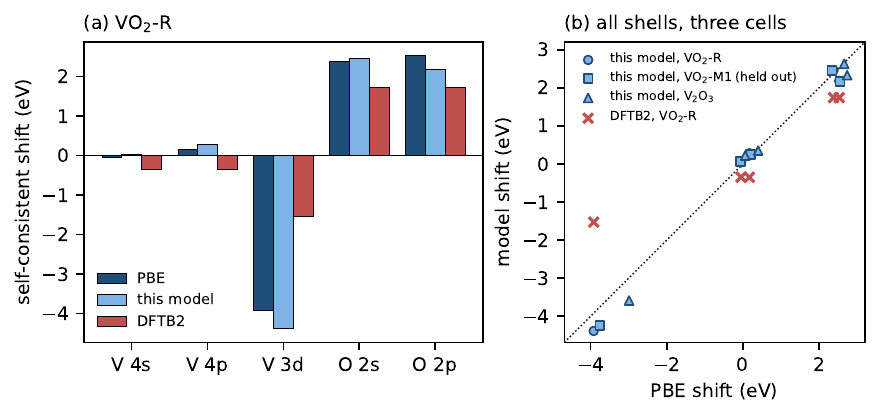}
\caption{The self-consistent shift of the on-site Hamiltonian for each shell is derived from a one-shift-per-shell fit, $\Delta H_{ij} \approx \tfrac12(\delta_i+\delta_j)S_{ij}$. (a) \ce{VO2}-R: a single-$\zeta$ span projection of PBE, the present model at $\theta^{(0)}$ employing the first radial of each shell, and the DFTB2 Mulliken--Slater charge model applied to a development version of the present reference. (b) The model is compared against PBE for all five shells of the three cells, with crosses indicating DFTB2 in rutile. The determination of $\theta^{(0)}$ did not include the M1 cell.}
\label{fig:diagnosis}
\end{figure}

We begin by investigating the modifications self-consistency introduces to the PBE Hamiltonian of \ce{VO2}-R, when projected onto a single-$\zeta$ atomic span. This change is fitted using one potential shift per shell, $\Delta H_{ij} \approx \tfrac12(\delta_i+\delta_j)S_{ij}$. The squared norm of the on-site diagonal change is captured to 99.6\% by the fit. Strong differences in the shifts appear between shells: V~3d moves down by \SI{3.92}{eV}, whereas V~4s and 4p barely move ($-0.05$ and $+0.17$~eV), and both O shells move up together ($+2.38$ and $+2.53$~eV). In contrast, the DFTB2 charge model, when applied to a development version of the present reference, yields shifts of only \SI{-1.53}{eV} for V~3d, $-0.35$~eV for V~4s and 4p, and $+1.74$~eV for the O shells (Figure~\ref{fig:diagnosis}a). Clearly, Slater-type clouds with widths set by free-atom Hubbard parameters cannot produce a potential for the compact 3d shell that differs so greatly from that of the diffuse 4s and 4p shells. The error lies in the spatial sampling of the induced potential, not in the magnitude of a Hubbard parameter. M1 and \ce{V2O3} exhibit the same pattern (V~3d $-3.76$ and $-2.99$~eV in PBE).

A large downward V~3d shift arises from the orbital-resolved response, whereas the V~4s and 4p shifts remain near zero. To match the PBE projection form, we limit the model's self-consistent change at $\theta^{(0)}$ to the first radial of each shell and apply the same fit. The fit accounts for 99.7--99.8\% of the squared norm of the on-site diagonal change, producing V~3d shifts of $-4.39$, $-4.25$ and \SI{-3.59}{eV} in rutile, M1 and \ce{V2O3}, whereas PBE gives $-3.92$, $-3.76$ and \SI{-2.99}{eV} (Figure~\ref{fig:diagnosis}b). The O shells shift upward by 2.17--\SI{2.63}{eV}, whereas the V~4s and 4p shells remain within \SI{0.35}{eV} of zero. M1 was held out, and no part of these shifts contributed to determining $\theta^{(0)}$. The V~3d shift, 0.47--\SI{0.60}{eV} more negative than in PBE, and the O~2p shift, 0.35--\SI{0.41}{eV} smaller, constitute the largest remaining differences. Although the two fits share the same form, their atomic functions are not identical: the PBE projection employs a single-$\zeta$ set from the development stage, while the model uses the first radials of the declared basis.

The reference deviates in the opposite direction. Projecting PBE's frozen-superposition Hamiltonian onto the DFTB2 basis yields a band-index separation $D$ of \SI{11.77}{eV} between the mean V~3d and O~2p band energies, whereas the value in the complete plane-wave basis is \SI{11.66}{eV}. Using two-centre Slater--Koster tables constructed in the standard manner from confined $3d^3 4s^2$ vanadium atoms, the DFTB2 reference yields \SI{7.69}{eV}.\cite{skprogs} While self-consistency reduces PBE's separation to \SI{6.41}{eV} in the DFTB2 basis (\SI{6.62}{eV} in the plane-wave basis), it decreases DFTB2's value only to \SI{7.02}{eV}. As a result, DFTB2 finishes approximately \SI{0.61}{eV} away from PBE in this centroid separation, even though its reference lies \SI{4.08}{eV} too low and its self-consistent shift amounts to one eighth of PBE's. It is possible for a model to match this separation for incorrect reasons. Matching this separation does not guarantee accurate charge redistribution or response. Both terms in this model approximate the PBE values: the rutile reference yields $D = \SI{11.76}{eV}$, and self-consistency decreases it by \SI{5.41}{eV} to \SI{6.35}{eV}, compared with \SI{5.04}{eV} for the plane-wave PBE bands (11.66 to \SI{6.62}{eV}). The reference separations for M1 and \ce{V2O3} are 11.70 and \SI{11.69}{eV}, respectively, whereas the self-consistent shifts amount to $-5.20$ and \SI{-4.84}{eV}.

\subsection{Reference energies at fixed geometry}
\begin{figure}
\centering
\includegraphics[width=0.6\linewidth]{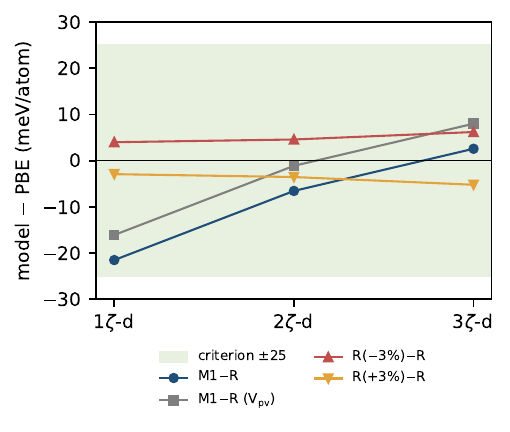}
\caption{The model minus PAW Harris--Foulkes reference-energy deviations are plotted as V~3d radials are added. The $\pm\SI{25}{meV/atom}$ criterion registered before the calculations is marked by the shaded band.}
\label{fig:gates}
\end{figure}

Three equal-composition pairs, namely dimerized M1 against rutile and rutile compressed and expanded by 3\% in volume, are compared in Table~\ref{tab:gates} via their model and plane-wave Harris--Foulkes reference-energy differences. The targets are not small in magnitude. At this level of theory, PBE positions M1 \SI{92.38}{meV/atom} above rutile, with the strained cells lying $+56.22$ and $-41.90$~meV/atom from it. For M1--R, the 3$\zeta$-d model produces a signed deviation of \SI{+2.59}{meV/atom}, while the volume pair shows $+6.21$ and $-5.20$~meV/atom, values that fall well within the $\pm\SI{25}{meV/atom}$ criterion registered before these values were computed (Figure~\ref{fig:gates}). When V~3p is included in the valence of the PAW reference, the M1--R deviation becomes $+8.02$~meV/atom. The 5.43~meV/atom spread quantifies how this difference responds to the two PAW datasets, including their different semicore treatment.

\begin{table}
\caption{For equal-composition pairs of \ce{VO2}, the reference-energy deviations are given as model minus PAW Harris--Foulkes (meV/atom). The PBE differences establish the scale. The 3$\zeta$-d model serves as the baseline, whereas the 1$\zeta$ and 2$\zeta$ variants keep one and two V~3d radials. We evaluate the Coulomb contribution to the core-level shifts in $C_0$ using the two-centre reduction (Section~\ref{sec:numerics}), with the original values provided in the Supporting Information.}
\label{tab:gates}
\begin{tabular}{lrrrr}
\toprule
pair & PBE $\Delta E$ & 1$\zeta$-d & 2$\zeta$-d & 3$\zeta$-d \\
\midrule
M1 $-$ R (V, 3p frozen) & $+92.38$ & $-21.50$ & $-6.54$ & $+2.59$ \\
M1 $-$ R (V\_pv) & $+86.95$ & $-16.08$ & $-1.11$ & $+8.02$ \\
R($-3\%$) $-$ R & $+56.22$ & $+4.01$ & $+4.60$ & $+6.21$ \\
R($+3\%$) $-$ R & $-41.90$ & $-2.90$ & $-3.53$ & $-5.20$ \\
\bottomrule
\end{tabular}
\end{table}

The outcome for the dimerized cell is governed by the d radials. The deviation stands at $-21.50$~meV/atom when only one 3d radial is included, a value close to the criterion; introducing the second radial shifts it to $-6.54$, while the third moves it to $+2.59$. The V~3d states are modified by dimerization. How the deviation depends on the d span reveals that a single confined radial fails to track this change. One radial already captures the volume pair ($+4.01/-2.90$), in contrast, and adding d radials moves its deviations slightly away from zero ($+6.21/-5.20$). Consequently, extra d radials do not improve the strained-cell deviations; we retain them as computed.

However, demonstrating that three radials suffice is still an open task. We test this by enlarging the basis along a sequence registered before its calculation, which includes a fourth d radial (S1), a diffuse d function (S2), O~2p breathing (S3), V~4s breathing (S4) and V~4p breathing (S5). At matched rank, the dimerized-cell deviation follows the sequence $+2.59 \to +4.65 \to +4.00 \to +3.79 \to +4.47 \to +4.73$~meV/atom, with the final two steps being $+0.69$ and $+0.26$. Consequently, the 3$\zeta$-d model sits \SI{2.1}{meV/atom} away from the most enriched span. Yet, because the enriched spans are nearly linearly dependent, the choice of read-out convention becomes significant. The maximum deviation from the candidate's value along the sequence reaches \SI{3.77}{meV/atom} at full rank (at S2), while 3.35 and \SI{3.44}{meV/atom} correspond to the two overlap thresholds. The terminal increment varies between $-1.34$ and $+0.53$~meV/atom across the four conventions. Convergence is not claimed in the sense of a completed uncertainty budget, because the radial and interpolation uncertainties of the enriched spans have not been measured. The finding is more specific: no tested enrichment, under any of the four read-outs, shifts the candidate's deviation by more than \SI{3.8}{meV/atom}, which is well within the criterion.

\subsection{Band alignment}
The band alignment is also determined by this same operator. The alignment of the two Hamiltonians is defined by the mean difference between PBE's local electrostatic potential and that of the model's superposed atoms in the region outside spheres of 2.5 (V) and 1.6~bohr (O) around the nuclei, where this difference exhibits a spatial standard deviation of about \SI{16}{meV}. In all four \ce{VO2} cells and in \ce{V2O3}, the mean O~2s and O~2p band energies of the 3$\zeta$-d model sit 0.035--\SI{0.042}{eV} above PBE's after this alignment (Table~\ref{tab:bands}). In the same five cells, the lowest V~3d band is positioned 0.143--\SI{0.163}{eV} above PBE's. The offset remains comparable across the tested valences and the dimerized cell, whereas it reaches \SI{0.66}{eV} in rutile when only one d radial is used. The d offset is lowered by each added d radial, decreasing further to \SI{0.114}{eV} at S5 in rutile without reordering any manifold. The remaining \SI{0.15}{eV} is not apportioned between the operator and residual basis and quadrature effects.

\begin{table}
\caption{The band alignment, defined as model minus PBE (eV), is reported for the mean O~2s and O~2p band energies and the lowest V~3d band. Unless noted, the values correspond to the 3$\zeta$-d model. The band read-out with the absolute $S^Q$ eigenvalue cutoff $10^{-4}$ is used for the offsets.}
\label{tab:bands}
\begin{tabular}{lrrr}
\toprule
cell & O~2s & O~2p & lowest V~3d \\
\midrule
\ce{VO2}-R, 1$\zeta$-d & $+0.043$ & $+0.059$ & $+0.659$ \\
\ce{VO2}-R, 2$\zeta$-d & $+0.041$ & $+0.042$ & $+0.304$ \\
\ce{VO2}-R & $+0.039$ & $+0.039$ & $+0.159$ \\
\ce{VO2}-R, S5 & $+0.023$ & $+0.027$ & $+0.114$ \\
\ce{VO2}-M1 & $+0.038$ & $+0.042$ & $+0.143$ \\
\ce{VO2}-R ($-3\%$) & $+0.041$ & $+0.041$ & $+0.163$ \\
\ce{VO2}-R ($+3\%$) & $+0.036$ & $+0.038$ & $+0.156$ \\
\ce{V2O3} & $+0.035$ & $+0.035$ & $+0.160$ \\
\bottomrule
\end{tabular}
\end{table}

\subsection{Charge response}
We obtain the coefficients $\theta^{(0)} = (s_\mathrm{Vsp}, s_\mathrm{Vd}, s_\mathrm O, \eta) = (0.597, 0.633, 0.934, 0)$ from the 22 matched responses of \ce{VO2}-R and \ce{V2O3} on the final operators. The unscreened-range boundary $\eta = 0$ is selected by the optimization, as $\chi^2$ increases monotonically from 23.5 to 23.8, 24.0, 24.5 and 26.0 at the sampled values $\eta = 0.005$, 0.01, 0.02 and \SI{0.05}{bohr^{-2}} after re-optimizing the three scale factors. The vanadium factors obtained from the fitting are comparable, with $s_\mathrm{Vsp} = 0.597$ and $s_\mathrm{Vd} = 0.633$. The weakest direction of the tolerance-weighted Jacobian is dominated by $s_\mathrm{Vsp}$ with an admixture of $\eta$, exhibiting a singular value of 2.6 compared to 86.9 for the strongest. Shell resolution is provided by the distinct orbital-density clouds and their charge-to-potential mapping, even when the scale factors are similar; whether the difference between the two vanadium factors matters was not tested by a constrained fit. The coefficients are not refitted at any later point in the text.

\begin{figure}
\centering
\includegraphics[width=0.6\linewidth]{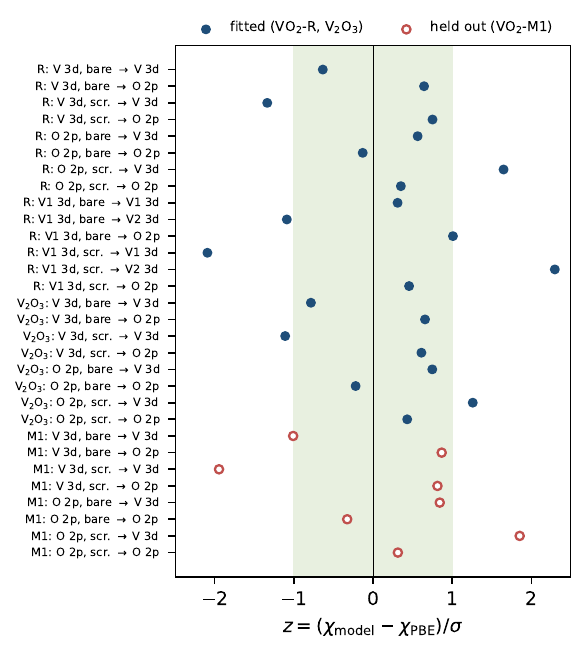}
\caption{The plot displays the normalized residual $z = (\chi_\mathrm{model}-\chi_\mathrm{PBE})/\sigma$ for the matched linear responses, where $\sigma = 0.10|\chi_\mathrm{PBE}| + \SI{0.005}{e/eV}$. The labels specify the cell, the field (shell, bare or screened) and the readout shell. The 22 responses used to determine $\theta^{(0)}$ appear as filled symbols; the 8 responses of the held-out M1 cell appear as open symbols. The band indicates $|z| \le 1$.}
\label{fig:response}
\end{figure}

For $\theta^{(0)}$, 14 of the 22 fitted responses match PBE within one tolerance and 20 within twice the tolerance, where $\chi^2 = 23.51$ over 22 serves as a comparison score rather than a significance test. The screened responses to the single-site rutile field, read on the perturbed V~3d shell ($z = -2.09$) and on its neighbour ($z = +2.30$), constitute the two components outside twice the tolerance; for the neighbour, the model gives \SI{+0.0182}{e/eV} against PBE's \SI{+0.0055}{e/eV}, with a tolerance dominated by the \SI{0.005}{e/eV} additive floor. M1, having been excluded from the fit, shows 5 of the 8 responses within one tolerance and all 8 within twice the tolerance ($\chi^2 = 10.55$, largest $|z| = 1.94$ for the screened V~3d self-response; Figure~\ref{fig:response}). The dimerized cell is described by coefficients determined on undimerized rutile and corundum. Systematic patterns persist. Across the three cells, 13 of the 15 bare responses exceed PBE's in magnitude by 2--17\%. The screened O~2p self-responses are 6--8\% smaller, whereas the O~2p readouts under V~3d fields are 19--25\% larger (Supporting Information, Section S8); among the eight screened V~3d readouts, seven are 17--40\% larger, and the remaining one, the small neighbour response in rutile, is 3.3 times PBE's. In 12 of the 15 channels, the screened-to-bare magnitude ratio is larger than PBE's, whereas it drops below that value in the three O~2p self-response channels. We leave this as an open item instead of absorbing it into the coefficients.

\subsection{Numerical control lies in the differences}
\label{sec:numerics} 
\begin{figure}
\centering
\includegraphics[width=0.6\linewidth]{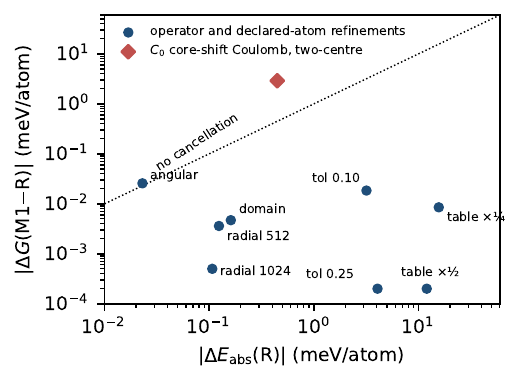}
\caption{For each tested refinement, the plot displays the change in the M1$-$R deviation against the change in the rutile cell's absolute reference energy. Cancellation in the difference is indicated by points located far below the dotted line. The diamond marks the two-centre evaluation replacing the sphere quadrature of the core-shift Coulomb contribution in $C_0$; this change does not cancel.}
\label{fig:channels}
\end{figure}

The relevant quantities are differences between two cells. Whether the numerical approximations cancel in those differences is the decisive question. Each refinement of the declared construction and its effect on a cell's absolute energy and on the deviations in Table~\ref{tab:gates} are shown in Table~\ref{tab:channels} and Figure~\ref{fig:channels}. Every cell's absolute energy shifts by about $+12.0$~meV/atom when the logarithmic table spacing of the declared atomic functions is halved. The shift is $+15.6$~meV/atom in the two cells tested over two halvings. Nevertheless, the M1--R deviation shifts by only $+0.0002$ and \SI{-0.0085}{meV/atom}, indicating that the cancellation exceeds 99.9\%.

The M1--R deviation shifts by $-0.026$, $-0.0005$ and $+0.005$, respectively, when the angular grid is doubled, the radial order is raised from 256 to 1024 points, and the integration domain is enlarged; altering the density matching tolerance moves it by $-0.018$ and \SI{+0.000}{meV/atom}. Only the angular grid refinement fails to cancel between the two cells, as rutile shifts by $+0.023$ and M1 by \SI{-0.003}{meV/atom}, making the change in their deviation, \SI{0.026}{meV/atom}, as large as the change in either cell. In the strained pair, the deviations shift by $-0.010$ and $-0.004$ upon angular doubling, by $-0.0088$ and $-0.0044$ with radial refinement, and by $+0.028$ and \SI{-0.019}{meV/atom} for table refinement. A table shift that is monotonic in volume ($+12.04$, $+12.01$ and \SI{+11.99}{meV/atom} for the compressed, reference and expanded cells) accompanies these last changes.

Using the original sphere quadrature of $C_0$, the same refinements altered M1--R by $+0.030$ and $+0.026$, while the strained deviations shifted by $+0.050$ and \SI{-0.039}{meV/atom}. The core-shift Coulomb quadrature error, detailed below, accounts for most of that change, as it relies on the tables itself. With the corrected evaluation, the operator-refinement results are observed sensitivities, not bounds; every value falls below \SI{0.03}{meV/atom}, compared to a criterion of 25. A rotation of the lattice representation of rutile shifts its absolute reference energy by \SI{0.69}{meV/atom}. Since we have not repeated this check for the other cells, it does not feature in the statement about deviations. The absolute reference energies remain provisional, because tested table refinements shift them by approximately \SI{15.6}{meV/atom}, and we do not quote any of them.

One component lies beyond these refinements, and it fails to cancel. The constant $C_0$ in eq~\ref{eq:eref} employs a distinct atom-centred quadrature. We integrated the first-order core-level shifts $\langle\chi_c|\sum_{(B,T)\ne(A,0)} v_{\mathrm{HN},B,T} + v_\mathrm{xc}[\rho_0] - v_\mathrm{xc}[\rho_A]|\chi_c\rangle$ on a sphere surrounding each atom, without any partition. Consequently, the sphere grid samples the Coulomb singularities of neighbouring nuclei located within it. We compute the Coulomb contribution using the two-centre reduction (analytic angular reduction, numerical radial integration) because the core density and each neighbour's potential are spherical, applying the identical formula as the pair term. The sphere evaluation accounted for the largest identified quadrature defect in the M1$-$R comparison. Substituting the Coulomb evaluation shifts the reference energies of R and M1 by $-0.4477$ and \SI{+2.4531}{meV/atom}, respectively, which raises the M1$-$R deviation by \SI{2.9008}{meV/atom}. The deviations of the compressed and expanded cells change by $-0.058$ and \SI{+0.059}{meV/atom}, as these cells shift by $-0.51$ and \SI{-0.39}{meV/atom}. The V~3p semicore shell accounts for almost all of the difference. The sphere evaluation underestimates the signed Coulomb contribution in M1 while overestimating it in rutile; as a result, the error does not cancel between the two cells (Figure~\ref{fig:channels}). The numerically corrected evaluation of the registered first-order frozen-core reference energy is reported in Table~\ref{tab:gates}. The original sphere-quadrature results are retained in the Supporting Information.

Sphere quadrature is retained for the exchange--correlation contribution to the core shifts and $\Delta X$. When the Coulomb contribution is evaluated by the two-centre reduction, angular doubling shifts the M1$-$R, compressed-R and expanded-R deviations by $+0.0176$, $+0.0845$ and \SI{+0.0256}{meV/atom}, whereas radial doubling shifts them by $-0.0378$, $+0.0112$ and \SI{+0.0070}{meV/atom}. Simultaneous doubling of both parameters alters the values by $-0.0575$, $+0.0049$ and \SI{+0.0392}{meV/atom}, a result that does not equal the sum of the two individual changes. The three deviations shift by $-0.006$, $+0.036$ and \SI{+0.071}{meV/atom} when the sphere radius is increased from 8 to 10~bohr with 200 radial points, a change that also moves all absolute energies by approximately \SI{-2.0}{meV/atom}. M1$-$R changes by \SI{+0.050}{meV/atom} when the number of images in the partition of $\Delta X$ is doubled, and by \SI{-0.0001}{meV/atom} when the grid on which the two-centre integrand is tabulated is halved. Holding the atomic functions fixed, a two- and fourfold refinement of the outer radial integration of the two-centre Coulomb integral changes the three deviations by $+0.0041$, $+0.0001$ and $+0.0009$, and by $+0.0024$, $+0.0002$ and \SI{+0.0006}{meV/atom}, respectively. Over the completed comparisons, these values serve as observed sensitivities, not bounds on the remaining quadrature error.

\begin{table}
\caption{For each tested refinement, the table gives (in meV/atom) the change in the rutile cell's absolute energy and $\Delta G$, the change in the model$-$PBE deviations of Table~\ref{tab:gates}. Refinements of the operator and declared atomic functions, including their effects on the corrected $C_0$, appear in the upper block, while the lower block details refinements of the quadrature of $C_0$.}
\label{tab:channels}
\small
\begin{tabular}{lrrrr}
\toprule
 & & \multicolumn{3}{c}{$\Delta G$} \\
\cmidrule(lr){3-5}
refinement & $\Delta E_\mathrm{abs}$(R) & M1$-$R & R($-3\%$)$-$R & R($+3\%$)$-$R \\
\midrule
angular $24{\times}48 \to 48{\times}96$ & $+0.023$ & $-0.026$ & $-0.010$ & $-0.004$ \\
radial $256 \to 512$ & $-0.124$ & $-0.004$ & --- & --- \\
radial $256 \to 1024$ & $-0.107$ & $-0.0005$ & $-0.0088$ & $-0.0044$ \\
table spacing halved\textsuperscript{\emph{c}} & $+12.008$ & $+0.000$ & $+0.028$ & $-0.019$ \\
table spacing halved twice\textsuperscript{\emph{c}} & $+15.650$ & $-0.0085$ & --- & --- \\
larger domain\textsuperscript{\emph{a}} & $-0.161$ & $+0.005$ & --- & --- \\
density tolerance $0.15\to0.10$\textsuperscript{\emph{c}} & $-3.185$ & $-0.018$ & $+0.002$ & $-0.002$ \\
density tolerance $0.15\to0.25$\textsuperscript{\emph{c}} & $+4.064$ & $+0.000$ & $-0.007$ & $+0.006$ \\
\midrule
$C_0$ core Coulomb: two-centre & $-0.448$ & $+2.901$ & $-0.058$ & $+0.059$ \\
$C_0$ sphere, angular $\times2$\textsuperscript{\emph{b}} & $+0.060$ & $+0.018$ & $+0.084$ & $+0.026$ \\
$C_0$ sphere, radial $\times2$\textsuperscript{\emph{b}} & $-0.557$ & $-0.038$ & $+0.011$ & $+0.007$ \\
$C_0$ sphere, radial and angular $\times2$\textsuperscript{\emph{b}} & $-0.461$ & $-0.058$ & $+0.005$ & $+0.039$ \\
$C_0$ sphere, radius $8\to10$~bohr\textsuperscript{\emph{b}} & $-2.016$ & $-0.006$ & $+0.036$ & $+0.071$ \\
$C_0$ partition, images $24\to48$\textsuperscript{\emph{b}} & $-0.002$ & $+0.050$ & --- & --- \\
$C_0$ two-centre integrand grid halved & $-0.003$ & $-0.0001$ & --- & --- \\
$C_0$ two-centre outer radial integral $\times2$ & $+0.0004$ & $+0.0041$ & $+0.0001$ & $+0.0009$ \\
$C_0$ two-centre outer radial integral $\times4$ & $-0.00003$ & $+0.0024$ & $+0.0002$ & $+0.0006$ \\
\bottomrule
\end{tabular}

\textsuperscript{\emph{a}}\,The inclusion radii are set to V $16.24\to19.5$ and O $10.95\to13.0$~bohr. \textsuperscript{\emph{b}}\,These tests refine the residual exchange--correlation component of the core-level shifts and $\Delta X$ with the two-centre Coulomb contribution held fixed (200 radial points for the radius test). \textsuperscript{\emph{c}}\,Evaluated with the two-centre Coulomb contribution, these refinements alter the declared atomic functions and hence $C_0$. ---: not computed.
\end{table}

%======================================================================
\section{Discussion}
%======================================================================
One property constrains the model's predictions: convexity. For ensemble density matrices with the given electron number, the spinless free energy $\Omega[P] = \mathrm{Tr}[P h^0] + \tfrac12 q[P]^\mathrm T\Gamma q[P] - k_\mathrm B T\,\mathcal S[P]$ at fixed geometry and electron number comprises a linear term and a convex quadratic term (the charges $q$ are linear in $P$ and $\Gamma$ is positive semidefinite). For any finite electronic temperature,\cite{Mermin1965} the expression further includes the negative of the strictly concave Fermi--Dirac entropy $\mathcal S$. Consequently, the free energy exhibits strict convexity, the self-consistent solution is unique, and the minimizer maintains every symmetry respected by the implemented operators and sampling. When $T = 0$, degeneracy at the Fermi level can eliminate uniqueness, allowing an iteration to halt at a symmetry-broken member of the degenerate set. Nevertheless, convexity guarantees that the symmetric average of any two solutions constitutes a solution sharing the same energy. Under these conditions, the model cannot spontaneously differentiate sites equivalent under the symmetries respected by the implemented operators and sampling, whereas inequivalent sites may carry different charges at fixed geometry. In the Magn\'eli phases, any spontaneous differentiation of symmetry-equivalent sites must therefore originate either from electron--lattice coupling---energy differences between competing dimer registries, with forces---or from an explicit spin or correlation term. This does not represent a deficiency of the reference or the response. It pins down the next quantity the model must compute and excludes such differentiation of symmetry-equivalent sites in a fixed-geometry calculation of the present model.

We present several unresolved matters and declare them as such. The model currently lacks forces, a spin block, and any test concerning unequal-composition energetics or Magn\'eli registry energies. The energetic comparisons reported here address Harris--Foulkes reference energies and do not establish the accuracy of self-consistent total-energy differences. We do not apportion the \SI{0.15}{eV} offset of the V~3d bands between operator and residual basis or quadrature effects. In 13 of 15 cases, the bare responses exceed PBE's by 2--17\%, while the screened-to-bare ratio surpasses PBE's in 12 of the 15 channels. A systematic discrepancy remains in the first-radial shell-shift comparison, where the V~3d shifts are 0.47--\SI{0.60}{eV} more negative than PBE's, while the O~2p shifts are 0.35--\SI{0.41}{eV} smaller. Computed on a construction that preceded the present one, the bare single-site responses of \ce{V4O7} were 35--40\% too large on the V atoms bonded to the perturbed site, and they await re-evaluation on the final operators. We present the enrichment sequence as a finite set of differences, rather than a converged limit accompanied by an error budget. The construction still has open targets for further development and validation, namely these discrepancies.

%======================================================================
\section{Conclusions}
%======================================================================
In the standard DFTB2 construction, the neutral two-centre reference lacks electron-volt environment terms, and the Mulliken charge model samples the induced potential with an incorrect spatial distribution; these two errors compensate in the band-index centroid separation between the V~3d and O~2p bands. When the reference is computed from superposed neutral PBE atoms with frozen-core orthogonality for the tested construction and cells, it matches PBE's closely without any adjustable term. Against the primary PAW reference at the stated settings, the associated Harris--Foulkes energy reproduces the tested plane-wave energy differences of \ce{VO2}, showing signed deviations of $+2.59$ (M1), $+6.21$ (3\% compression) and \SI{-5.20}{meV/atom} (3\% expansion). We determine three screening factors for an orbital-resolved, shell-screened charge response from rutile and corundum, placing all eight tested M1 response components within twice the stated tolerance; across the three cells, the first-radial shell-shift fits yield V~3d shifts 0.47--\SI{0.60}{eV} more negative than PBE's and O~2p shifts 0.35--\SI{0.41}{eV} smaller. The Magn\'eli charge and dimer patterns become the next target through electron--lattice coupling, given that the spinless model is convex at fixed geometry. Other early transition-metal oxides can be treated with the same construction without change in form.

\section*{Data and Software Availability}
The model parameters are available at \url{https://github.com/anerjy/vo-tb-parameters}.

\begin{acknowledgement}
The author gratefully acknowledges the long-term and steadfast support from the Department of Physics, the Faculty of Arts and Sciences, and the University of Puerto Rico.

Under the author's direction, an AI agent was used to write and run the construction and analysis scripts and draft the manuscript.
\end{acknowledgement}

\begin{suppinfo}
The declared atomic functions, including their WKB continuation, supports, and matching rule, are constructed and their sensitivities assessed; quadrature certificates are provided; full enrichment ladders are presented under all read-out conventions; per-band offsets are listed; model and PBE values for all 30 response components are given; and a numerical-channel table for every cell is included.
\end{suppinfo}

\bibliography{refs}

\end{document}